\documentclass[journal]{IEEEtran}
\usepackage{amsmath}
\usepackage{amssymb}
\usepackage{amsfonts}
\usepackage{graphicx}
\usepackage{epsfig}
\usepackage{subfigure}
\usepackage{psfrag}
\usepackage{cite}
\usepackage{latexsym}
\usepackage{url}
\usepackage{color}
\usepackage{algorithm}
\usepackage{algorithmic}

\PassOptionsToPackage{bookmarks={false}}{hyperref}

\newtheorem{proposition}{\underline{Proposition}}[section]

\newcommand{\mv}[1]{\mbox{\boldmath{$ #1 $}}}

\begin{document}
\title{Cognitive Wireless Power Transfer in the Presence of Reactive Primary Communication User}
\author{Tianxin Feng, {\it Student Member, IEEE}, Ganggang Ma, {\it Student Member, IEEE}, and Jie Xu, {\it Member, IEEE},
\thanks{Manuscript received September 30, 2018; revised February 2, 2019; accepted March 18, 2019. (Corresponding author: Jie Xu.)

T. Feng and G. Ma are with the School of Information Engineering, Guangdong University of Technology, Guangzhou 510006, China (e-mail:~ftx.gdut@gmail.com,~gangma.gdut@gmail.com).

J. Xu is with the School of Information Engineering, Guangdong University of Technology, Guangzhou 510006, China, and also with the National Mobile Communications Research Laboratory, Southeast University, Nanjing 211189, China (e-mail: jiexu@gdut.edu.cn).}}

\maketitle

\begin{abstract}
This paper studies a cognitive or secondary multi-antenna wireless power transfer (WPT) system over a multi-carrier channel, which shares the same spectrum with a primary wireless information transfer (WIT) system that employs adaptive water-filling power allocation. By controlling the transmit energy beamforming over sub-carriers (SCs), the secondary energy transmitter (S-ET) can directly charge the secondary energy receiver (S-ER), even purposely interfere with the primary WIT system, such that the primary information transmitter (P-IT) can reactively adjust its power allocation (based on water-filling) to facilitate the S-ER's energy harvesting. We investigate how the secondary WPT system can exploit the primary WIT system's reactive power allocation, for improving the wireless energy harvesting performance. In particular, our objective is to maximize the total energy received at the S-ER from both the S-ET and the P-IT, by optimizing the S-ET's energy beamforming over SCs, subject to its maximum transmit power constraint, and the maximum interference power constraint imposed at the primary information receiver (P-IR) to protect the primary WIT. Although the formulated problem is non-convex and difficult to be optimally solved in general, we propose an efficient algorithm to obtain a high-quality solution by employing the Lagrange dual method together with a one-dimensional search. We also present two benchmark energy beamforming designs based on the zero-forcing (ZF) and maximum-ratio-transmission (MRT) principles, respectively, as well as the conventional design without considering the primary WIT system's reaction. Numerical results show that our proposed design leads to significantly improved energy harvesting performance at the S-ER, as compared to these benchmark schemes.
\end{abstract}

\section{Introduction}
\IEEEPARstart{R}{adio} frequency (RF) signals enabled wireless power transfer (WPT) has been recognized as a promising technique to provide controllable and convenient energy supply for low-power wireless devices such as sensors and RF identification (RFID) tags. WPT can help these devices avoid frequent battery replacement to achieve self-sustainable operation, and thus has attracted a lot of recent research interests for future Internet of things (IoT) networks with massive connectivity \cite{1,2,Xia}. Besides, WPT has also found abundant applications by integrating with other emerging techniques such as wireless communications and mobile (edge) computing. For example, simultaneous wireless information and power transfer (SWIPT) \cite{12,15,16,Costa} and wireless powered communication networks (WPCN) \cite{20,22,222} unify WPT and wireless communications in a joint design, which exploit the dual use of RF signals for WPT and wireless information transfer (WIT) over the same and opposite transmission directions, respectively. Furthermore, wireless powered edge computing \cite{you,222222,Bi} employs WPT to achieve self-sustainable computation for wireless devices at the network edge, which can use the harvested wireless energy for both local computing and task offloading.

Due to the severe signal propagation loss over distance, however, the wide implementation of WPT is hindered by the low end-to-end energy transfer efficiency. In the literature, there have been various techniques proposed to tackle this issue, such as waveform optimization \cite{511,111}, adaptive power allocation \cite{Zeng,51}, and energy beamforming \cite{12,16,Huang}. For example, the energy transmitter (ET) can properly design the transmit signal waveforms based on the non-linear model of RF-to-direct current (DC) circuits at the energy receiver (ER), in order to enhance the harvested DC energy for efficiently charging the battery at the ER \cite{511,111}. Under given transmit waveforms, the authors in \cite{Zeng} and \cite{51} proposed wideband WPT (or SWIPT) designs, in which the ET can adaptively adjust its power allocation over sub-carriers (SCs) to exploit the frequency diversity for efficient energy transfer. Furthermore, multi-antenna energy beamforming is another promising technique, which enables the ET to adjust the transmit energy beamforming directions towards the intended ERs, thus significantly improving the energy transfer efficiency \cite{12,16,Huang}. To fully reap the benefit of adaptive power allocation and energy beamforming, it is crucial for the ET to acquire the channel state information (CSI) with the ERs. In practice, such information can be obtained by the ERs via implementing the energy measurement feedback \cite{3,4,5} or the reverse-link channel training based on the channel reciprocity between the forward and reverse links \cite{11}.
By exploiting these techniques, various start-up companies (such as TransferFi, Energous, and Powercast\footnote{Please refer to \url{http://www.transferfi.com}, \url{http://www.energous.com}, and \url{http://www.powercastco.com} for more details, respectively.}) in the industry have demonstrated commercial WPT products for efficiently charging sensors and IoT devices at a long distance (e.g., more than ten meters (m)).

Conventional WPT systems are operated over dedicated frequency bands. This, however, is difficult to be implemented in practice due to the scarcity of spectrum resources. Recently, cognitive radio (CR) with spectrum sharing has attracted a lot of attentions in the wireless communications society, which enables the secondary wireless communication systems to share the same spectrum resources that are originally allocated to primary wireless communication systems (see, e.g., \cite{Goldsmith} and the references therein). In general, there are three approaches in CR networks, namely interweave, underlay and overlay. In the interweave approach, secondary users detect the dynamic spectrum holes over time for opportunistic access \cite{hay}. In the underlay approach, the secondary users are allowed to access the spectrum if the interference caused to the primary users is below a given interference temperature (IT) threshold \cite{sad}. In the overlay approach, the secondary users employ sophisticated signal processing and coding techniques to help the primary system in order to access some spectrum for its own transmission \cite{g}. In this paper, we focus our study on the underlay approach with the IT technique. Inspired by its great success in wireless communications, it is expected that the spectrum sharing technique is also an efficient solution to improve the WPT performance, by allowing the WPT system to act as a cognitive or secondary wireless system, which can access the frequency bands that are originally allocated to primary wireless communication or WIT systems.

The spectrum sharing between secondary WPT and primary WIT systems introduces new technical challenges to be dealt with. In particular, the secondary WPT system results in an interesting one-way interference towards the primary WIT system \cite{20}. On one hand, the transmitted energy signals by the secondary ET (S-ET) induce {\it harmful} interference to the primary information receiver (P-IR); but on the other hand, the transmitted information signals by the primary information transmitter (P-IT) can be additionally exploited as a {\it beneficial} energy source that can be harvested by the secondary ER (S-ER). Under the one-way interference, the authors in \cite{32} considered a new information-helping WPT system, in which the S-ET helps the P-IT in the wireless information transmission, in order to access the spectrum for delivering wireless power to the S-ER. Furthermore, the authors in \cite{322} considered the spectrum sharing between a secondary multiple-input multiple-output (MIMO) WPT system and a coexisting point-to-point MIMO WIT system, in which the S-ET optimizes the energy beamforming to maximize the harvested energy at the S-ER, while minimizing the interference power towards the P-IR. In addition, there have been various prior works investigating the spectrum sharing for SWIPT \cite{29,26,25} and WPCN \cite{23,31,30,24,27,28}.

Nevertheless, all the previous works assumed that the primary WIT system adopts fixed transmission strategies by using constant transmit power and beamforming vectors. In practice, however, the adaptive transmission has been widely adopted in modern wireless communication systems, which can adaptively adjust the transmit power allocation and transmit beamforming according to the wireless channel and interference conditions. In this case, the secondary WPT system can exploit the reaction of the primary WIT system in transmission adaptation, to increase the energy harvesting performance. Intuitively, the S-ET can purposely interfere with the WIT system to change the interference conditions, such that the WIT system can accordingly adapt its transmit power and beamforming directions towards the S-ER, thus facilitating the wireless energy harvesting. How to optimally exploit such an effect for maximizing the WPT performance in secondary WPT systems is an interesting problem that has not been studied yet, thus motivating our investigation in this work. The main results of this paper are summarized as follows.
\begin{itemize}
\item This paper focuses on a multi-antenna secondary WPT system over multi-carrier channels, which shares the same spectrum band with a primary WIT system that employs adaptive water-filling power allocation. This may correspond to a practical scenario in next-generation self-sustainable IoT networks with coexisting WPT and WIT systems over the same frequency bands. To improve the energy harvesting performance of the energy-hungry node (the S-ER), the secondary WPT system designs the transmit energy beamforming over SCs to not only directly charge the S-ER, but also control the one-way interference towards the P-IR, such that the P-IT can reactively adjust its power allocation (based on water-filling) to facilitate the wireless energy harvesting at the S-ER.
\item Our objective is to maximize the received RF power at the S-ER from both the S-ET and the P-IT, by optimizing the S-ET's energy beamforming over SCs, subject to the maximum transmit power constraints at the S-ET and the IT threshold constraint imposed at the P-IR to protect the primary WIT. Although the formulated problem is non-convex and difficult to be optimally solved in general, we propose an efficient algorithm to obtain a high-quality solution by employing the Lagrange dual method together with a one-dimensional search.
\item We also present two benchmark energy beamforming designs based on the zero-forcing (ZF) and maximum-ratio-transmission (MRT) principles, respectively, as well as the conventional design without considering the primary WIT system's reaction. Numerical results show that our proposed design significantly outperforms these benchmark schemes, especially when the IT threshold and/or the number of transmit antennas at S-ET become large.
\end{itemize}


The remainder of this paper is organized as follows. Section \ref{sec:system} introduces the system model and formulates the problem of our interest. Section \ref{sec:Solution} proposes an efficient algorithm to solve this problem by employing the Lagrange dual method together with a one-dimensional search. Section \ref{low} presents two benchmark energy beamforming designs based on the ZF and MRT principles, respectively, as well as the conventional design without considering the primary WIT system's reaction. Section \ref{sec:Results} provides numerical results to show the performance of our proposed design. Section \ref{sec:Conclusion} concludes this paper.

\emph{Notations}: Letters in bold denote the vectors (lower case) or matrices (upper case). For a square matrix $\mv{H}$, ${\rm tr}(\mv{H})$ refers to its trace, while $\mv{H}\succeq \mv{0}$ means that $\mv{H}$ is positive semi-definite. For an arbitrary-size matrix $\mv{A}$, $\|\mv{A}\|_F$, $\mv{A}^H$, and ${\rm{rank}}(\mv{A})$ denote its Frobenius norm, conjugate transpose, and rank, respectively. $\mathbb{C}^{M\times N}$ refers to the space of $M\times N$ complex matrices. Furthermore, $\mv{I}$ and $\mv{0}$ denote the identity matrix and the all-zero matrix with appropriate dimensions.

\section{System Model and Problem Formulation}\label{sec:system}

\begin{figure}[h]
\centering
\includegraphics[width=7cm]{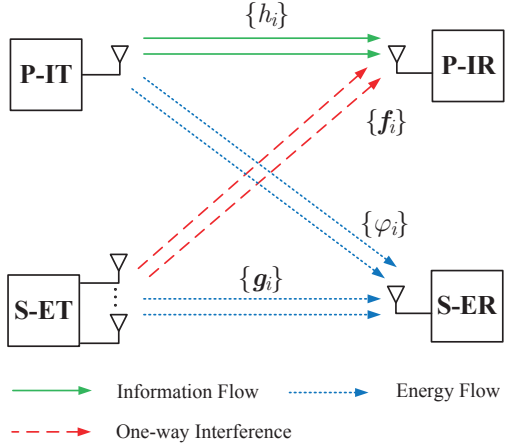}
\caption{A spectrum sharing scenario between a primary point-to-point WIT system and a secondary point-to-point multi-antenna WPT system.}	\label{fig:1}
\end{figure}

As shown in Fig. \ref{fig:1}, we consider the spectrum sharing between a primary WIT system and a secondary WPT system, where the P-IT transmits information signals to the P-IR while the S-ET delivers wireless energy to the S-ER.\footnote{Notice that the secondary WPT and primary WIT systems are separate. In this case, our considered setup is different from the SWIPT system with WPT and WIT integrated in the same system.} The P-IT, P-IR, and S-ER are each equipped with one single antenna, and the S-ET is equipped with $M>1$ antennas for efficient WPT. Both the primary WIT and secondary WPT systems operate over the same spectrum band, which consists of $N$ SCs each with identical bandwidth, with $\mathcal{N} \triangleq \{1, . . . , N\}$ denoting the set of the $N$ SCs.

We consider a time-invariant frequency-selective block fading channel model, in which the wireless channels remain unchanged over each time slot of our interest. Let $h_i$ and $\varphi_i$ denote the channel power gains from the P-IT to the P-IR and the S-ER over SC $i \in \mathcal N$, respectively. Let $\mv{g}_i \in\mathbb{C}^{M\times 1}$ and $\mv{f}_i\in\mathbb{C}^{M\times 1}$ denote the channel vectors from the S-ET to the S-ER and the P-IR over SC $i \in \mathcal N$, respectively. It is assumed that the S-ET perfectly knows the global CSI of $\{h_i\}$, $\{\varphi_i\}$, $\{\mv{g}_i\}$, and $\{\mv {f}_i\}$, for the purpose of characterizing the fundamental performance limits of the secondary WPT system.

First, we consider the information signal transmission over the primary WIT system. At each SC $i \in \mathcal N$, let $s_i$ denote the transmitted information signal by the P-IT, and $x_i$ denote the transmitted energy signal by the S-ET. We assume that $s_i$ and $x_i$ are independent circularly symmetric complex Gaussian (CSCG) random variables with zero mean and unit variance, i.e., $s_i \sim \mathcal{CN}(0,1)$ and $x_i \sim \mathcal{CN}(0,1)$. Notice that the S-ET can implement more sophisticated transmit waveform design for $x_i$ (instead of using simple CSCG random variables) to further enhance the WPT performance \cite{511,111}. However, how to analyze the interference towards the P-IR under such designs is a challenging task, and therefore, we consider the CSCG signaling for both primary WIT and secondary WPT for the ease of analysis. Let $\mv {\omega}_i \in \mathbb{C}^{M\times 1}$ denote the transmit energy beamforming vector at the S-ET for SC $i$, and $P_i$ denote the transmit power at the P-IT. In this case, the received signal at the P-IR over SC $i$ is expressed as
\begin{align}
y_i =  \sqrt{h_i P_i} s_i + \mv{f}^H_i \mv {\omega}_i x_i + n_i,\ i\in\mathcal N,
\end{align}
where $n_i$ denotes the noise at the P-IR receiver, which is a CSCG random variable with zero mean and variance $\sigma^2$. Accordingly, the achievable rate over each SC $i$ at the P-IR (in bps/Hz) is
\begin{align}
R_i ={\log _2}\left(1 + \frac{h_iP_i}{|{\mv{f}^H_i}\mv{\omega}_i|^2+\sigma^2}\right),\ i\in\mathcal N.\label{equ:ECS2}
\end{align}
Suppose that the P-IT adopts the water-filling power allocation to maximize the sum-rate throughput $\sum_{i\in\mathcal N} R_i$ of the primary WIT system, by optimizing the transmit power allocation $\{P_i\}$ over SCs under any given energy beamforming $\{\mv{\omega}_i\}$, subject to a maximum sum transmit power $P_{\rm sum}$. Therefore, the sum-rate maximization problem for the primary WIT system is given by
\begin{align}
\max\limits_{\{P_i\}} ~&\sum\limits_{i \in\mathcal N}{\log _2}\left(1 + \frac{h_iP_i}{|{\mv{f}^H_i}\mv{\omega}_i|^2+\sigma^2}\right)\label{equ:ECSq3}\\
\mathrm{s.t.}~&\sum\limits_{i \in\mathcal N} P_i\leq{P_{\mathrm{sum}}}\label{equ:ECSq2108}\\
&P_i\geq0,\ \forall i\in \mathcal{N}.
\end{align}
It is well established that the optimal solution to problem (\ref{equ:ECSq3}) corresponds to the celebrated water-filling power allocation (see, e.g., \cite{37}), i.e.,
\begin{align}
{P^*_i}(\mv{\omega}_i,\lambda)=\bigg(\lambda-\frac{|{\mv{f}^H_i}\mv{\omega}_i|^2+\sigma^2}{h_i}\bigg)^+,\label{equ:ECS4}
\end{align}
where $x^+\triangleq {\max}(x,0)$, and $\lambda>0$ denotes the water level that can be obtained based on the following equation:
\begin{align}
\sum\limits_{i \in\mathcal N} P_i^*(\mv{\omega}_i,\lambda)=P_{\mathrm{sum}}.\label{equ:ECS5}
\end{align}
It is evident that as long as the transmit energy beamforming vectors $\mv{\omega}_i$'s are changed, the primary WIT system should correspondingly adapt the water-filling power allocation based on (\ref{equ:ECS5}). This motivates us to design $\{\mv{\omega}_i\}$ to control the energy transferred from the P-IT in the primary WIT system to the S-ER, thus improving the wireless energy harvesting performance, as will be detailed later.

Next, we consider the wireless energy harvesting at the secondary WPT system. Due to the broadcast nature of RF signals, the S-ER can harvest the wireless energy carried by both the energy signals $\{x_i\}$ sent from the S-ET and the information signals $\{s_i\}$ sent from the P-IT. Accordingly, the received RF power by the S-ER is expressed as
\begin{align}
Q_{\rm{ER}}(\{\mv{\omega}_i\},\lambda) =\sum\limits_{i\in\mathcal N} \left({P_i^*(\mv{\omega}_i,\lambda){\varphi_i}+{|{\mv{g}^H_i}\mv{\omega}_i|}^2}\right)\label{equ:ECS1}.
\end{align}
It is worth noting that based on (\ref{equ:ECS1}), the received power by the S-ER consists of two terms that come from the P-IT and the S-ET, respectively. On one hand, the S-ET can design the energy beamforming vectors towards the S-ER to maximize the directly transferred power $\sum_{i\in\mathcal N} \|\mv{g}^H_i\mv{\omega}_i\|^2$; on the other hand, by properly interfering with the P-IR, the S-ET can also affect the P-IT's power allocation, thus changing the transferred power $\sum_{i\in\mathcal N} P_i^*(\mv{\omega}_i,\lambda) \varphi_i$  from the P-IT to the S-ER. Therefore, for maximizing the total energy received at the S-ER, the design of $\{\mv{\omega}_i\}$ at the S-ET needs to properly balance the tradeoff between the above two terms. It is also worth noting that in order to charge the battery at the S-ER, the received RF signals need to be converted into DC signals via rectifiers \cite{11}, and the RF-to-DC energy conversion process is generally non-linear, especially when the input RF power comes significantly small and large. In particular, at the medium to large input RF power regimes (e.g., larger than 0.1 mW \cite{371}), a sigmoidal function-based RF-to-DC energy conversion model has been proposed in \cite{non-linearModel}, which is obtained by using curve fitting based on practical measurement results. At the low RF input power regime (e.g., less than 0.1 mW \cite{371}), another non-linear RF-to-DC energy conversion model based on the rectifier's circuit characteristics has been proposed in \cite{511} to facilitate the transmit waveform optimization. Notice that for both models, the received DC power is a monotonically non-decreasing function with respect to the input RF power, under the fixed transmit waveform with Gaussian signaling. In this case, maximizing the received DC power is equivalent to maximizing the input RF power. Therefore, in this paper we directly focus on the received RF power in (8) before the RF-to-DC energy conversion. This is not only for analytical tractability, but also practically relevant.

Suppose that the S-ET is subject to a maximum sum transmit power constraint $Q_{\rm sum}$, and a set of peak transmit power constraints $Q_{\rm peak}$ each for one SC. We thus have
\begin{align}
 &\sum\limits_{i\in\mathcal N} {\|\mv{\omega}_i\|}^2\leq Q_{\mathrm{sum}},\label{09202056}\\
 &\|\mv{\omega}_i\|^2 \leq Q_{\mathrm{peak}},\ \forall i\in \mathcal{N}.\label{09202057}
\end{align}
Furthermore, in order to protect the primary WIT's transmission with certain QoS guaranteed, we impose the IT constraint at the P-IR, such that the sum interference power from the S-ET to the P-IR over all the SCs does not exceed an IT threshold $\Gamma$. Thus, we have
\begin{align}
\sum\limits_{i\in\mathcal N} |{\mv{f}^H_i}\mv{\omega}_i|^2 \le \Gamma.\label{09202058}
\end{align}

Our objective is to maximize the total received power $Q_{\rm{ER}}(\{\mv{\omega}_i\},\lambda)$ in (\ref{equ:ECS1}) from both the S-ET and the P-IT, by optimizing the S-ET's transmit energy beamforming $\{\mv{\omega}_i\}$ over the $N$ SCs, subject to the S-ET's power constraints in (\ref{09202056}) and (\ref{09202057}), the P-IR's IT constraint in (\ref{09202058}), and the constraint in (\ref{equ:ECS5}) due to the water-filling power allocation adopted at the P-IT. The sum received power maximization problem is thus formulated as
\begin{align}
 {\mathtt{(P1)}}:\max\limits_{\{\mv{\omega}_i\},\lambda>0}&\sum\limits_{i\in\mathcal N} \left({P_i^*(\mv{\omega}_i,\lambda){\varphi_i}+|{\mv{g}^H_i}\mv{\omega}_i|^2}\right)\label{equ:ECS11}\\
 \mathrm{s.t.}~ &(\ref{equ:ECS5}),~(\ref{09202056}),~(\ref{09202057}), ~\text {and}~ (\ref{09202058}).\nonumber
\end{align}
It is observed that the objective function of problem ${\mathtt{(P1)}}$ is non-concave,
and the equality constraint in (\ref{equ:ECS5}) is not affine. Therefore, problem ${\mathtt{(P1)}}$ is generally non-convex, and thus very difficult to be solved optimally.

\section{Proposed Solution to Problem (${\mathtt{P1}}$)}\label{sec:Solution}
 In this section, we present an efficient algorithm to solve problem ${\mathtt{(P1)}}$, in which we first optimize $\{\mv{\omega}_i\}$ under any given $\lambda$ by using the Lagrange dual method, and then use a one-dimensional search to find the optimal $\lambda$. In the following, we first obtain the regime of $\lambda$ in order for problem ${\mathtt{(P1)}}$ to be feasible, and then we focus on solving problem ${\mathtt{(P1)}}$ under given $\lambda$ within this regime by optimizing $\{\mv{\omega}_i\}$ only.

\subsection{Finding Feasible Regime of $\lambda$}\label{sec:SolutionA}
In this subsection, we obtain the lower and upper bounds of $\lambda$ in order for problem (${\mathtt{P1}}$) to be feasible, which are denoted by $\lambda_{\rm min}$ and $\lambda_{\rm max}$, respectively. First, it is observed from (\ref{equ:ECS5}) that in order for the equality to hold, the water level $\lambda$ generally increases as the interference power $\{|\mv{f}_i^H \mv{\omega}_i|^2\}$ increases. Therefore, it follows that $\lambda_{\rm min}$ is attained when the interference power is minimized to be zero, i.e., $\mv{\omega}_i=\mathbf{0},\ \forall i\in\mathcal N$. In other words, we have
\begin{align}
\sum_{i\in\mathcal N}\left(\lambda_{\rm min} - \frac{\sigma^2}{h_i}\right) = P_\mathrm{sum},
\end{align}
based on which $\lambda_{\rm min}$ can be easily found via a simple bisection search.

Next, it remains to obtain the upper bound $\lambda_{\rm max}$. Similarly, it is evident that $\lambda_{\rm max}$ is attained when the interference power from the S-ET to the P-IR is maximized. In this case, the S-ET should align the transmit energy beamforming vectors towards the P-IR by setting
\begin{align}
\mv{\omega}_i=\sqrt{Q_i}\frac{\mv{f}_i}{\|\mv{f}_i\|},\ \forall i\in\mathcal {N},\label{09211422}
\end{align}
 where $Q_i$'s denote the S-ET's transmit power allocated over different SCs, $\forall i\in\mathcal N$, which are variables to be determined. To obtain $\{Q_i\}$ for finding $\lambda_{\rm max}$,  we consider the following feasibility problem under given $\lambda$, by substituting (\ref{09211422}) into problem (${\mathtt{P1}}$).
\begin{align}
{\mathtt{(P2):}} ~\mathop\mathtt{find} &~~\{Q_i\}\label{equ:ECS6}\\
\mathrm{s.t.}&~\sum\limits_{i\in\mathcal N} \|\mv{f}_i\|^2Q_i \le \Gamma\label{equ:ECS08301718}\\
&~\sum\limits_{i\in\mathcal N} Q_i \le Q_{\mathrm{sum}}\label{equ:ECS08301719}\\
&~0\leq Q_i \le Q_{\mathrm{peak}},\forall i\in \mathcal{N}\label{equ:ECS08301720}\\
&~\sum\limits_{i\in\mathcal N} \bigg(\lambda-\frac{\|\mv{f}_i\|^2Q_i+\sigma^2}{h_i}\bigg)^+= P_{\mathrm{sum}}.\label{equ:ECS08301721}
\end{align}
Notice that problem ${\mathtt{(P2)}}$ is feasible when $\lambda_{\mathrm{min}} \le \lambda \le \lambda_{\mathrm{max}}$, but infeasible when $\lambda > \lambda_{\mathrm{max}}$. Therefore, we can find $\lambda_{\mathrm{max}}$ by checking the feasibility of (${\mathtt{P2}}$) under any given $\lambda > \lambda_{\mathrm{min}}$, and then adopt a bisection search over $\lambda$. In the following, we focus on solving problem (${\mathtt{P2}}$) under given $\lambda > \lambda_{\mathrm{min}}$.

 It is observed that problem ${\mathtt{(P2)}}$ is still non-convex as the equality constraint in (\ref{equ:ECS08301721}) is not affine. However, it is worth noting that the time sharing condition defined in \cite{34} holds for problem ${\mathtt{(P2)}}$ when the number of SCs $N$ goes to infinity,\footnote{In our simulations, it is observed that even with a finite number of SCs $N=64$, the strong duality still holds between problem $\mathtt{(P2)}$ and its dual problem.} and therefore, the strong duality holds between problem ${\mathtt{(P2)}}$ and its dual problem in this case \cite{34}. As a result, we use the Lagrange dual method to solve problem ${\mathtt{(P2)}}$.

Let $\eta\geq 0$, $\mu\geq 0$, and $\theta$ denote the dual variables associated with the constraints in (\ref{equ:ECS08301718}), (\ref{equ:ECS08301719}), and (\ref{equ:ECS08301721}), respectively. Then, the partial Lagrangian of problem ${\mathtt{(P2)}}$ is given as
\begin{align}
&\mathcal{L}(\{Q_i\},\eta,\mu,\theta)=\nonumber\\
&-\eta\bigg(\sum\limits_{i\in\mathcal N} \|\mv{f}_i\|^2Q_i-\Gamma\bigg)-\mu\bigg(\sum\limits_{i\in\mathcal N} Q_i-Q_{\mathrm{sum}}\bigg)\nonumber\\
&-\theta\bigg(\sum\limits_{i\in\mathcal N} \bigg(\lambda-\frac{\|\mv{f}_i\|^2Q_i+\sigma^2}{h_i}\bigg)^+-P_{\mathrm{sum}}\bigg).
\end{align}
The dual function is expressed as
\begin{align}
D(\eta,\mu,\theta)=\max\limits_{\{Q_i\}} ~&\mathcal{L}\left(\{Q_i\},\eta,\mu,\theta\right)\label{equ:ECS10}\\
\mathrm{s.t.}~&0\leq Q_i \le Q_{\mathrm{peak}}.\nonumber
\end{align}
Accordingly, the dual problem becomes
\begin{gather}
\min \limits_{\eta\geq0,\mu\geq0,\theta} ~D(\eta,\mu,\theta).\label{equ:ECS20}
\end{gather}

As strong duality holds between problem ${\mathtt{(P2)}}$ and its dual problem (\ref{equ:ECS20}), we solve problem ${\mathtt{(P2)}}$ by equivalently solving the dual problem (\ref{equ:ECS20}).
In the following, we first obtain $D(\eta,\mu,\theta)$ by solving problem (\ref{equ:ECS10}), and then update the dual variables $\eta$, $\mu$, and $\theta$ to minimize $D(\eta,\mu,\theta)$.

Under given $\eta$, $\mu$, and $\theta$, problem (\ref{equ:ECS10}) can be decomposed into $N$ subproblems as follows by ignoring the constant term $\eta \Gamma + \mu Q_{\mathrm{sum}} + \theta P_{\mathrm{sum}}$, in which each subproblem corresponds to one SC $i\in\mathcal N$.
\begin{align}
\max\limits_{Q_i}~&-\eta \|\mv{f}_i\|^2Q_i-\mu Q_i-\theta\bigg(\lambda-\frac{\|\mv{f}_i\|^2Q_i+\sigma^2}{h_i}\bigg)^+\label{equ:ECSb}\\
\mathrm{s.t.}~&0\leq Q_i \le Q_{\mathrm{peak}}.\label{equ:06010006}
\end{align}
It is observed that problem (\ref{equ:ECSb}) is non-linear and non-convex due to the term $\left(\lambda-\frac{\|\mv{f}_i\|^2Q_i+\sigma^2}{h_i}\right)^+$. Accordingly, we solve problem (\ref{equ:ECSb}) by considering the following three cases based on the sign of $\lambda-\frac{\|\mv{f}_i\|^2Q_i+\sigma^2}{h_i}$.

Case (i): Consider that $\frac{ h_i \lambda - \sigma^2}{\|{\mv{f}_i}\|^2}<0$. As $Q_i \ge 0$ holds in \eqref{equ:06010006}, we always have $Q_i> \frac{ h_i \lambda - \sigma^2}{\|{\mv{f}_i}\|^2}$ or equivalently $\lambda-\frac{\|\mv{f}_i\|^2Q_i+\sigma^2}{h_i} < 0$. Accordingly, problem ($\ref{equ:ECSb}$) can be expressed as
\begin{align}
\max\limits_{Q_i} ~&-(\eta\|\mv{f}_i\|^2+\mu){Q_i}\label{equ:ECS0526}\\
\mathrm{s.t.}~&(\ref{equ:06010006}).\nonumber
\end{align}
Problem (\ref{equ:ECS0526}) is a linear program (LP) with the coefficient $-(\eta\|\mv{f}_i\|^2+\mu) < 0$. Therefore, the optimal solution to problem (\ref{equ:ECS0526}) is $Q^{(1)}_i=0$ and the corresponding optimal value is $0$.

Case (ii): consider that $\frac{ h_i \lambda - \sigma^2}{\|{\mv{f}_i}\|^2}> Q_{\mathrm{peak}}$. As $Q_i \le Q_{\mathrm{peak}}$ holds in \eqref{equ:06010006}, we always have $Q_i< \frac{ h_i \lambda - \sigma^2}{\|{\mv{f}_i}\|^2}$ or equivalently $\lambda-\frac{\|\mv{f}_i\|^2Q_i+\sigma^2}{h_i} > 0$. Therefore, problem ($\ref{equ:ECSb}$) is equivalent to
\begin{align}
\max\limits_{Q_i}~&(-\eta\|\mv{f}_i\|^2-\mu+\frac{\theta\|\mv{f}_i\|^2}{h_i})Q_i-\theta (\lambda-\frac{\sigma^2}{h_i})\label{equ:05301432}\\
\mathrm{s.t.}~&(\ref{equ:06010006}).\nonumber
\end{align}
Problem (\ref{equ:05301432}) is an LP, whose optimal solution can be obtained by considering the following two cases based on the sign of the coefficient $-\eta\|\mv{f}_i\|^2-\mu+\frac{\theta\|\mv{f}_i\|^2}{h_i}$. If $-\eta\|\mv{f}_i\|^2-\mu+\frac{\theta\|\mv{f}_i\|^2}{h_i}>0$, the optimal solution is ${Q^{(2)}_i}=Q_{\mathrm{peak}}$, and the corresponding optimal value is $\left(-\eta\|\mv{f}_i\|^2-\mu+\frac{\theta\|\mv{f}_i\|^2}{h_i}\right)Q_{\mathrm{peak}}-\theta \left(\lambda-\frac{\sigma^2}{h_i}\right)$. Otherwise, the optimal solution is ${Q^{(2)}_i}=0$, and the corresponding optimal value is $-\theta \left(\lambda-\frac{\sigma^2}{h_i}\right)$.

Case (iii): consider that $0\leq\frac{ h_i \lambda - \sigma^2}{\|{\mv{f}_i}\|^2}\leq Q_{\mathrm{peak}}$, for which we need to consider the following two sub-cases.
\begin{itemize}
\item In the first subcase with $0\leq Q_i\leq \frac{ h_i \lambda - \sigma^2}{\|{\mv{f}_i}\|^2}$, problem ($\ref{equ:ECSb}$) is equivalent to maximizing the objective function in (\ref{equ:05301432}) subject to $0\leq Q_i\leq \frac{ h_i \lambda - \sigma^2}{\|{\mv{f}_i}\|^2}$. Similar to case (ii), if $-\eta\|\mv{f}_i\|^2-\mu+\frac{\theta\|\mv{f}_i\|^2}{h_i}>0$, the optimal solution is ${Q^{(3)}_i}=\frac{ h_i \lambda - \sigma^2}{\|{\mv{f}_i}\|^2}$, and the optimal value is $\Psi_1=\left(-\eta\|\mv{f}_i\|^2-\mu\right)\frac{ h_i \lambda - \sigma^2}{\|{\mv{f}_i}\|^2}$; otherwise, the optimal solution is ${Q^{(3)}_i}=0$, and the optimal value is $\Psi_1=-\theta \left(\lambda-\frac{\sigma^2}{h_i}\right)$.
\item In the second subcase with $\frac{ h_i \lambda - \sigma^2}{\|{\mv{f}_i}\|^2}\leq Q_i \leq Q_{\mathrm{peak}}$, problem ($\ref{equ:ECSb}$) is equivalent to maximizing the objective function in (\ref{equ:ECS0526}) subject to $\frac{ h_i \lambda - \sigma^2}{\|{\mv{f}_i}\|^2}\leq Q_i \leq Q_{\mathrm{peak}}$. In this subcase, the optimal solution is ${Q^{(4)}_i}=\frac{ h_i \lambda - \sigma^2}{\|{\mv{f}_i}\|^2}$, and the optimal value is $\Psi_2=\left(-\eta\|\mv{f}_i\|^2-\mu\right)\frac{ h_i \lambda - \sigma^2}{\|{\mv{f}_i}\|^2}$.
\end{itemize}
By comparing the optimal values $\Psi_1$ and $\Psi_2$ for the two subcases, we can obtain the optimal solution to problem \eqref{equ:ECSb} in case (iii) as the one corresponding to the larger optimal value.

By combining the three cases (i), (ii), and (iii), the optimal solution to problem (\ref{equ:ECSb}) is summarized as
\begin{align}
&Q_i^{\star}(\eta,\mu,\theta)=\nonumber\\
&\left\{\begin{array}{ll}
\frac{ h_i \lambda - \sigma^2}{\|{\mv{f}_i}\|^2},  &{\rm{if}}\ 0\leq\frac{ h_i \lambda - \sigma^2}{\|{\mv{f}_i}\|^2}\leq Q_{\mathrm{peak}}\ \text{and}\ \\
&(\Psi_1<\Psi_2\ \text{or}\ -\eta\|\mv{f}_i\|^2-\mu+\frac{\theta\|\mv{f}_i\|^2}{h_i}>0),\\
Q_{\rm peak},&{\rm if}\ \frac{ h_i \lambda - \sigma^2}{\|{\mv{f}_i}\|^2}> Q_{\mathrm{peak}}\ \text{and}\\ &-\eta\|\mv{f}_i\|^2-\mu+\frac{\theta\|\mv{f}_i\|^2}{h_i}>0,\\
0, &{\rm otherwise.}
 \end{array} \right.\label{Q}
\end{align}
By substituting $Q_i^{\star}(\eta,\mu,\theta)$'s, $\forall i\in\mathcal N$, into \eqref{equ:ECS10}, the dual function $D(\eta,\mu,\theta)$ is obtained.

Next, based on the obtained $D(\eta,\mu,\theta)$, we solve the dual problem (\ref{equ:ECS20}). As the dual function $D(\eta,\mu,\theta)$ is always convex but generally non-differentiable, problem (\ref{equ:ECS20}) is a convex optimization problem. Therefore, we solve this problem by subgradient based methods such as the ellipsoid method \cite{36}, based on the fact that the subgradient of $D(\eta,\mu,\theta)$ is given as
\begin{align}
&\mv{s}(\eta,\mu,\theta)=\nonumber\\
&\bigg[\Gamma-\sum\limits_{i\in\mathcal N} \|\mv{f}_i\|^2Q_i^*(\eta,\mu,\theta),~Q_{\mathrm{sum}}-\sum\limits_{i\in\mathcal N} Q_i^*(\eta,\mu,\theta),\nonumber\\
&P_{\mathrm{sum}}-\sum\limits_{i\in\mathcal N} \bigg(\lambda-\frac{\|{\mv{f}_i}\|^2Q_i^*(\eta,\mu,\theta)+\sigma^2}{h_i}\bigg)^+\bigg].\label{subg}
\end{align}

More specifically, we have the following proposition that helps design terminate conditions for the iterations in the ellipsoid method.
\begin{proposition}\label{proposition:2}
Problem ${\mathtt{(P2)}}$ is infeasible if and only if there exist $\eta\geq0$, $\mu\geq0$, and $\theta$, such that $D(\eta,\mu,\theta)<0$.
\end{proposition}
\begin{IEEEproof}
This proposition follows due to the strong duality between the feasibility problem ${\mathtt{(P2)}}$ and the dual problem (\ref{equ:ECS20}). The detailed proof is similar to that in \cite[Appendix B]{duan}, and thus is omitted for brevity.
\end{IEEEproof}

 Based on Proposition \ref{proposition:2}, it is evident that when problem ${\mathtt{(P2)}}$ is infeasible, the iteration in the ellipsoid method will lead to a set of dual variables $\eta$, $\mu,$ and $\theta$ with $D(\eta,\mu,\theta)<0$. In this case, the iteration will terminate, and it is ensured that problem ${\mathtt{(P2)}}$ is infeasible and $\lambda>\lambda_{\rm max}$ follows. Otherwise, if problem ${\mathtt{(P2)}}$ is feasible, the iteration will converge to an optimal dual value of zero, due to the strong duality between the primal problem ${\mathtt{(P2)}}$ and its dual problem (\ref{equ:ECS20}). In this case, it is ensured that $\lambda\leq\lambda_{\rm max}$.

 In summary, the detailed algorithm to solve the feasibility problem ${\mathtt{(P2)}}$ under any given $\lambda>\lambda_{\rm min}$ is presented in Algorithm 1.

\begin{algorithm}[htbp]
    \caption{for solving problem ${\mathtt{(P2)}}$}
    \begin{algorithmic}[1]
    \REQUIRE
    Set the initial values of $\eta \ge 0$, $\mu \ge 0$, and $\theta$, as well as an initial ellipsoid that contains the optimal dual solution of $\eta$, $\mu$, and $\theta$;
    \REPEAT
    \STATE Obtain the optimal solution to problem (\ref{equ:ECS10}) under given $\eta$, $\mu$, and $\theta$ as $Q_i^{\star}(\eta,\mu,\theta)$'s, $\forall i\in\mathcal N$, in (\ref{Q});
    \STATE If $D(\eta,\mu,\theta)<0$, then output that problem ${\mathtt{(P2)}}$ is infeasible, exit the loop, and the algorithm ends; otherwise, go to the next step;
    \STATE Use the ellipsoid method to update $\eta$, $\mu$, and $\theta$, based on the subgradient $\mv{s}(\eta,\mu,\theta)$ in (\ref{subg});
    \UNTIL convergence;
    \STATE Output that problem ${\mathtt{(P2)}}$ is feasible.
    \end{algorithmic}
\end{algorithm}

Finally, by implementing Algorithm 1 for solving problem ${\mathtt{(P2)}}$ under any given $\lambda$, together with a bisection search over $\lambda>\lambda_{\mathrm{min}}$, the upper bound $\lambda_\mathrm{max}$ can be found.

{\it Complexity.} The computational complexity of Algorithm 1 is dominated by the ellipsoid method in steps $1-5$. In particular, during each iteration, the computational complexity is dominated by step 2, which is of order $MN$. Let $\varepsilon_1$ denote the accuracy for ellipsoid method, and thus the computational complexity of an ellipsoid method is $\mathcal{O}(\log(1/\varepsilon_1))$ \cite{36}. Accordingly, under any given $\lambda$, Algorithm 1 for solving the feasibility problem (P2) has a complexity of $\mathcal{O}(MN\log(1/\varepsilon_1))$. Furthermore, let $\varepsilon_2$ denote the accuracy for bisection over $\lambda$. Therefore, the computational complexity of finding $\lambda_\mathrm{max}$ is $\mathcal{O}(MN\log(1/\varepsilon_1)\log(1/\varepsilon_2))$.

\subsection{ Optimizing \{$\mv{\omega}_i$\} for Problem ${\mathtt{(P1)}}$ under Given $\lambda\in [{\lambda}_{\rm min},{\lambda}_{\rm max}]$}
In this subsection, we optimize $\{\mv{\omega}_i\}$ for problem ${\mathtt{(P1)}}$ under given $\lambda\in [{\lambda}_{\rm min},{\lambda}_{\rm max}]$. In this case, the optimization problem is explicitly expressed as
\begin{align}
{\mathtt{(P3)}}:\max\limits_{\{\mv{\omega}_i\}}&\sum\limits_{i\in\mathcal N} \left({P_i^*(\mv{\omega}_i){\varphi_i}+|{\mv{g}^H_i}\mv{\omega}_i|^2}\right)\\
 \mathrm{s.t.}~ &(\ref{equ:ECS5}),~(\ref{09202056}),~(\ref{09202057}), ~\text {and}~ (\ref{09202058}).\nonumber
\end{align}
Problem $\mathtt{(P3)}$ is still non-convex, as the objective function is not concave, and the equality constraint in (\ref{equ:ECS5}) is not affine. Nevertheless, similarly as for problem ${\mathtt{(P2)}}$, the time sharing condition holds for this problem when $N \to \infty $. Therefore, the strong duality holds between problem ${\mathtt{(P3)}}$ and its dual problem in this case \cite{34}.\footnote{In our simulations, it is also observed that even with a finite number of SCs $N=64$, the strong duality still holds between problem ${\mathtt{(P3)}}$ and its dual problem.} As a result, we use the Lagrange dual method to solve problem ${\mathtt{(P3)}}$.

Let $\eta_1\geq0$, $\mu_1\geq0$, and $\theta_1$ denote the dual variables associated with the constraints in (\ref{equ:ECS5}), (\ref{09202056}), and (\ref{09202058}), respectively. Then the partial Lagrangian of problem ${\mathtt{(P3)}}$ is expressed as
\begin{align}
&\mathcal{L}_1(\{\mv{\omega}_i\},\eta_1,\mu_1,\theta_1)=\nonumber\\
&\sum\limits_{i\in\mathcal N} \bigg({\bigg(\lambda-\frac{|{\mv{f}^H_i}\mv{\omega}_i|^2+\sigma^2}{h_i}\bigg)^+{\varphi_i}+|{\mv{g}^H_i}\mv{\omega}_i|^2}\bigg)\nonumber\\
&-\eta_1\bigg(\sum\limits_{i\in\mathcal N} |{\mv{f}^H_i}\mv{\omega}_i|^2-\Gamma\bigg)-\mu_1\bigg(\sum\limits_{i\in\mathcal N} \|\mv{\omega}_i\|^2-Q_{\mathrm{sum}}\bigg)\nonumber\\
&-\theta_1\bigg(\sum\limits_{i\in\mathcal N} \bigg(\lambda-\frac{|{\mv{f}^H_i}\mv{\omega}_i|^2+\sigma^2}{h_i}\bigg)^+-P_{\mathrm{sum}}\bigg).
\end{align}
The dual function is
\begin{align}
D_1\left(\eta_1,\mu_1,\theta_1\right)=\max\limits_{\{\mv{\omega}_i\}}~& ~\mathcal{L}_1\left(\{\mv{\omega}_i\},\eta_1,\mu_1,\theta_1\right)\label{equ:ECS16}\\
\mathrm{s.t.}~&\|\mv{\omega}_i\|^2 \leq Q_{\mathrm{peak}},\forall i\in \mathcal{N}.\nonumber
\end{align}
Accordingly, the dual problem is given as
\begin{gather}
\min \limits_{\eta_1\geq0,~\mu_1\geq0,~\theta_1} D_1\left(\eta_1,\mu_1,\theta_1\right).\label{equ:ECSF}
\end{gather}

As strong duality holds between problem (P3) and its dual problem (\ref{equ:ECSF}), we solve problem ${\mathtt{(P3)}}$ by equivalently solving the dual problem (\ref{equ:ECSF}). In the following, we first obtain $D_1(\eta_1,\mu_1,\theta_1)$ by solving problem (\ref{equ:ECS16}) under given $\eta_1\ge 0$, $\mu_1\ge 0$, and $\theta_1$, and then update the dual variables $\eta_1$, $\mu_1$, and $\theta_1$ to minimize $D_1(\eta_1,\mu_1,\theta_1)$. Under given $\eta_1\ge 0$, $\mu_1\ge 0$, and $\theta_1$, problem (\ref{equ:ECS16}) can be decomposed into $N$ subproblems as follows by ignoring the constant term $\eta_1\Gamma+\mu_1Q_{\mathrm{sum}}+\theta_1 P_{\mathrm{sum}}$, in which each subproblem corresponds to one SC $i\in\mathcal N$.

\begin{align}
\mathtt{(P4)}: \nonumber\\
\max\limits_{\mv{\omega}_i}~~& {\bigg(\lambda-\frac{|{\mv{f}^H_i}\mv{\omega}_i|^2+\sigma^2}{h_i}\bigg)^+{\varphi_i}+|{\mv{g}^H_i}\mv{\omega}_i|^2}
-\eta_1 |{\mv{f}^H_i}\mv{\omega}_i|^2 \nonumber\\
&-\mu_1\|\mv{\omega}_i\|^2-\theta_1\bigg(\lambda-\frac{|{\mv{f}^H_i}\mv{\omega}_i|^2+\sigma^2}{h_i}\bigg)^+\label{equ:05311112}\\
\mathrm{s.t.}~~&\|\mv{\omega}_i\|^2 \leq Q_{\mathrm{peak}}.\label{equ:5311112}
\end{align}

In order to solve problem ${\mathtt{(P4)}}$, we consider the following two cases with $\lambda-\frac{|{\mv{f}^H_i}\mv{\omega}_i|^2+\sigma^2}{h_i} \le 0$ and $\lambda-\frac{|{\mv{f}^H_i}\mv{\omega}_i|^2+\sigma^2}{h_i} \ge 0$, respectively.
\begin{itemize}

\item First, we consider the case with $\lambda-\frac{|{\mv{f}^H_i}\mv{\omega}_i|^2+\sigma^2}{h_i} \le 0$, or equivalently, $|{\mv{f}^H_i}\mv{\omega}_i|^2\geq h_i \lambda - \sigma^2$. In this case, problem (P4) can be re-expressed as
\begin{align}
\mathtt{(P4.1)}:\max\limits_{\mv{\omega}_i}~& {|{\mv{g}^H_i}\mv{\omega}_i|^2}-\eta_1 |{\mv{f}^H_i}\mv{\omega}_i|^2-\mu_1 \|\mv{\omega}_i\|^2 \nonumber\\
\mathrm{s.t.}~&|{\mv{f}^H_i}\mv{\omega}_i|^2\geq h_i \lambda - \sigma^2 \label{09282340}\\
&\|\mv{\omega}_i\|^2 \leq Q_{\mathrm{peak}}.\label{Q_peak:1}
\end{align}
Notice that if $Q_{\mathrm{peak}} < (h_i \lambda - \sigma^2)/\|\mv{f}\|^2$, then we cannot find $\mv\omega_i$ to satisfy constraints (\ref{09282340}) and \eqref{Q_peak:1} at the same time, and problem ${\mathtt{(P4.1)}}$ is infeasible. In this case, we set the objective value of problem $\mathtt{(P4.1)}$ as $\Psi_3=-\infty$ for notational convenience in this case. Otherwise, notice that problem ${\mathtt{(P4.1)}}$ is a quadratically constrained quadratic program (QCQP) that is generally non-convex. We use the semi-definite relaxation (SDR) technique to obtain its optimal solution. Let $\mv{G}_i={\mv{g}_i}{\mv{g}^H_i}$, $\mv{F}_i=\mv{f}_i{\mv{f}^H_i}$, and $\mv{W}_i=\mv{\omega}_i{\mv{\omega}^H_i}$. Thus, we equivalently re-express problem ${\mathtt{(P4.1)}}$ as
\begin{align}
\mathtt{(P4.2)}:\max \limits_{\mv{W}_i}~&{\rm{tr}}((\mv{G}_i-\eta_1 \mv{F}_i-\mu_1 \mv{I})\mv{W}_i)\label{equ:05311140}\\
\mathtt{s.t.}~&{\rm{tr}}(\mv{F}_i\mv{W}_i)\geq h_i \lambda - \sigma^2 \label{equ:0601000}\\
&{\rm{tr}}(\mv{W}_i)\leq Q_{\mathrm{peak}}\label{equ:05311524}\\
&{\rm{rank}}(\mv{W}_i)\leq 1\label{equ:ECSrank}\\
&\mv{W}_i\succeq \mathbf{0}.\label{equ:05311525}
\end{align}
Problem $\mathtt{(P4.2)}$ is still non-convex, as the rank constraint in (\ref{equ:ECSrank}) is not convex. Fortunately, by dropping constraint (\ref{equ:ECSrank}), this problem becomes a semi-definite program (SDP), referred to as ${\mathtt{(P4.3)}}$, which can be optimally solved via the standard convex optimization techniques, e.g. by using CVX \cite{37}. Notice that as the SDP ${\mathtt{(P4.3)}}$ has two linear constraints, it always has a rank-one solution \cite{35}. As a result, the SDR is actually tight. Let ${\mv W}^{(1)}_i$ denote the rank-one solution to problem ${\mathtt{(P4.3)}}$ and thus ${\mathtt{(P4.2)}}$. Then we obtain the optimal solution to ${\mathtt{(P4.3)}}$ as $\mv{\omega}_i^{(1)}$ by performing the eigenvalue decomposition (EVD) of ${\mv W}^{(1)}_i$. Accordingly, the objective value in this case is obtained $\Psi_3=\mv{\omega}_i^{(1)H}(\mv{G}_i-\eta_1 \mv{F}_i-\mu_1 \mv{I})\mv{\omega}_i^{(1)}$.

\item Next, consider the other case when $\lambda-\frac{|{\mv{f}^H_i}\mv{\omega}_i|^2+\sigma^2}{h_i}\ge0$, or equivalently,
 $|{\mv{f}^H_i}\mv{\omega}_i|^2\le h_i \lambda - \sigma^2$. In this case, problem ${\mathtt{(P4)}}$ is expressed as the following non-convex QCQP:
\begin{align}
\mathtt{(P4.4)}:\max\limits_{\mv{\omega}_i} &~ {|{\mv{g}^H_i}\mv{\omega}_i|^2}+\bigg(-\eta_1+\frac{\theta_1-\varphi_i}{h_i}\bigg) |{\mv{f}^H_i}\mv{\omega}_i|^2\nonumber\\
&-\mu_1 \|\mv{\omega}_i\|^2+(\varphi_i-\theta_1) (\lambda-\sigma^2/h_i)\\
\mathrm{s.t.}&~|{\mv{f}^H_i}\mv{\omega}_i|^2\leq h_i \lambda - \sigma^2\\
&~\|\mv{\omega}_i\|^2 \leq Q_{\mathrm{peak}}.
\end{align}
Notice that when $h_i \lambda - \sigma^2 < 0$, problem $\mathtt{(P4.4)}$ is infeasible and we set its optimal value as $\Psi_4=-\infty$ in this case. Therefore, we only focus on problem $\mathtt{(P4.4)}$ with $h_i \lambda - \sigma^2\ge 0$, for which the SDR is used to obtain the optimal solution. By introducing $\mv{G}_i={\mv{g}_i}{\mv{g}^H_i}$, $\mv{F}_i=\mv{f}_i{\mv{f}^H_i}$, and $\mv{W}_i=\mv{\omega}_i{\mv{\omega}^H_i}$, problem ${\mathtt{(P4.4)}}$ is equivalently expressed as
\begin{align}
\mathtt{(P4.5)}:\nonumber\\
\max\limits_{\mv{W}_i}&~{\rm{tr}}\bigg(\bigg(\mv{G}_i+\left(-\eta_1+\frac{\theta_1-\varphi_i}{h_i}\right) \mv{F}_i-\mu_1 \mv{I}\bigg)\mv{W}_i\bigg)\nonumber\\
&+(\varphi_i-\theta_1) (\lambda-\sigma^2/h_i)\label{equ:05311515}\\
\mathrm{s.t.}&~{\rm{tr}}(\mv{F}_i\mv{W}_i)\leq h_i \lambda - \sigma^2\label{equ:09140124}\\
&~(\ref{equ:05311524}), ~(\ref{equ:ECSrank}), ~\text{and} ~(\ref{equ:05311525}).\nonumber
\end{align}

Similarly as for problem ${\mathtt{(P4.2)}}$, we can obtain the optimal solution to this QCQP problem ${\mathtt{(P4.5)}}$ via SDR, for which the detailed procedure is omitted for brevity. Let $\mv{\omega}_i^{(2)}$ denote the optimal solution to problem ${\mathtt{(P4.4)}}$. Accordingly, the objective value in this case is obtained $\Psi_4=\mv{\omega}_i^{(2)H}(\mv{G}_i+(-\eta_1+\frac{\theta_1-\varphi_i}{h_i}) \mv{F}_i-\mu_1 \mv{I})\mv{\omega}_i^{(2)}+(\varphi_i-\theta_1) (\lambda-\frac{\sigma^2}{h_i})$.
\end{itemize}

Now, we compare $\Psi_3$ and $\Psi_4$ in the above two cases to obtain the optimal solution to problem $\mathtt{(P4)}$. If $\Psi_3 \ge \Psi_4$, then we have $\mv\omega_i^{**}(\eta_1,\mu_1,\theta_1) = \mv{\omega}_i^{(1)}$. Otherwise, we have $\mv\omega_i^{**}(\eta_1,\mu_1,\theta_1) = \mv{\omega}_i^{(2)}$.
\begin{algorithm}[htbp]
    \caption{for solving problem $\mathtt{(P3)}$}
    \begin{algorithmic}[1]
    \REQUIRE
    Set the initial values of $\eta_1 \ge 0$, $\mu_1  \ge 0$, and $\theta_1 $, as well as an initial ellipsoid that contains the optimal dual solution of $\eta_1$, $\mu_1$, and $\theta_1$;
    \REPEAT
    \STATE Obtain the optimal solution to problem (\ref{equ:ECS16}) as $\mv\omega_i^{**}(\eta_1,\mu_1,\theta_1)$'s, $\forall i\in\mathcal N$, under given $\eta_1 $, $\mu_1 $, and $\theta_1 $;
    \STATE Use the ellipsoid method to update $\eta_1$, $\mu_1$, and $\theta_1$, based on the subgradient $\mv{s}_1(\eta_1,\mu_1,\theta_1)$ in (\ref{09252355});
    \UNTIL convergence;
    \STATE Set $\eta_1^{*} \gets \eta_1$, $\mu_1^{*} \gets \mu_1$, and $\theta_1^{*} \gets \theta_1$. Accordingly, $\{\mv{\omega}_i^{**}(\eta_1^*,\mu_1^*,\theta_1^*)\}$ becomes the optimal solution to problem (\ref{equ:ECSF}).
    \end{algorithmic}
\end{algorithm}

Next, with $\{\mv\omega_i^{**}(\eta_1,\mu_1,\theta_1)\}$ obtained at hand, we solve the dual problem (\ref{equ:ECSF}). As the dual function $D_1(\eta_1,\mu_1,\theta_1)$ is always convex but generally non-differentiable, problem (\ref{equ:ECSF}) is a convex optimization problem. Therefore, we solve this problem via the ellipsoid method \cite{36}, by using the fact that the subgradient of $D_1(\eta_1,\mu_1,\theta_1)$ is given as
\begin{align}
&\mv{s}_1(\eta_1,\mu_1,\theta_1)=\nonumber\\
&\bigg[\Gamma-\sum\limits_{i\in\mathcal N} |{\mv{f}^H_i}\mv{\omega}_i^{**}(\eta_1,\mu_1,\theta_1)|^2,\nonumber\\
&Q_{\mathrm{sum}}-\sum\limits_{i\in\mathcal N} \|\mv{\omega}_i^{**}(\eta_1,\mu_1,\theta_1)\|^2,\nonumber\\
&P_{\mathrm{sum}}-\sum\limits_{i\in\mathcal N} \bigg(\lambda-\frac{|{\mv{f}^H_i}\mv{\omega}_i^{**}(\eta_1,\mu_1,\theta_1)|^2+\sigma^2}{h_i}\bigg)^+\bigg].\label{09252355}
\end{align}
Let $\eta_1^*$, $\mu_1^*$, and $\theta_1^*$ denote the optimal dual solution to problem (\ref{equ:ECSF}). By substituting them into $\mv\omega_i^{**}(\eta_1,\mu_1,\theta_1)$ above, we finally obtain the optimal primal solution to problem $\mathtt{(P3)}$. In summary, we present the detailed algorithm to solve (\ref{equ:ECS16}) as Algorithm 2.

By implementing Algorithm 2 for solving problem ${\mathtt{(P3)}}$, together with a one-dimensional search over $\lambda \in [{\lambda}_{\rm min},{\lambda}_{\rm max}]$, problem $\mathtt{(P1)}$ is finally solved.

{\it Complexity.} Let $\varepsilon_3$ denote the accuracy for solving SDP via interior point method, and $\varepsilon_4$ denote the accuracy for exhaust search. In step 2 in Algorithm 2, solving the SDP requires the complexity of $\mathcal{O}(\log ({1/\varepsilon_3}))$ \cite{com} and repeats $N$ times. By combining the ellipsoid method with complexity $\mathcal{O}(\log ({1/\varepsilon_1}))$, matrix multiplication with complexity $\mathcal{O}(M^3)$, and the exhaustive search over $\lambda$ with complexity $\mathcal{O}(1/\varepsilon_4)$, the time complexity of solving problem ${\mathtt{(P3)}}$ is $\mathcal{O}(M^3N\log ({1/\varepsilon_1})\log ({1/\varepsilon_3})/\varepsilon_4)$. After considering Algorithm 1 and 2, also the bisection and one-dimensional search, we can figure out the time complexity of solving problem ${\mathtt{(P1)}}$ is $\mathcal{O}(MN\log(1/\varepsilon_1)\log(1/\varepsilon_2)+M^3N\log ({1/\varepsilon_1})\log ({1/\varepsilon_3})/\varepsilon_4)$.


\section{Benchmark Schemes}\label{low}
In this section, we present several benchmark schemes for solving problem $\mathtt{(P1)}$. First, we propose two energy beamforming solutions  based on the ZF and MRT principles, respectively. Next, we consider the conventional energy beamforming design without considering the primary WIT system's reaction.
\subsection{ZF-Based Energy Beamforming}

In this subsection, we design the energy beamforming at the S-ET based on the ZF principle, which can perfectly cancel the interference from the S-ET to the P-IR, i.e., $|{\mv{f}^H_i}\mv{\omega}_i|^2 = 0,\forall i\in \mathcal{N}$. In this case, problem $\mathtt{(P1)}$ is simplified as
\begin{align}
 \mathtt{(P5)}:\max\limits_{\{\mv{\omega}_i\}} ~&\sum\limits_{i\in\mathcal N} \left|{\mv{g}^H_i}\mv{\omega}_i\right|^2\\
 \mathrm{s.t.} &\sum\limits_{i\in\mathcal N} \|\mv{\omega}_i\|^2 \leq Q_{\mathrm{sum}}\label{09251519}\\
   &\|\mv{\omega}_i\|^2 \leq Q_{\mathrm{peak}},\forall i\in \mathcal{N}\label{09251510}\\
   &|{\mv{f}^H_i}\mv{\omega}_i|^2 =0 ,\forall i\in \mathcal{N}.\label{09242346}
\end{align}
Based on (\ref{09242346}), the S-ET should set each transmit beamforming vector $\mv{\omega}_i$ within the null space of $\mv{f}_i$, $\forall i\in\mathcal N$. In order to maximize $\|{\mv{g}^H_i}\mv{\omega}_i\|$ in this case, we have ${\mv{\omega}^{\text{ZF}}_i}=\sqrt{Q_i} \frac{(\mv{I}-\mv{f}_i\mv{f}_i^H)^{-1}\mv{g}_i}{\|(\mv{I}-\mv{f}_i\mv{f}_i^H)^{-1}\mv{g}_i\|}$, where $Q_i \ge 0$ is the transmit power at each SC $i\in\mathcal N$ to be decided.
 By substituting ${\mv{\omega}^{\text{ZF}}_i}$'s in problem ${\mathtt{(P5)}}$, problem ${\mathtt{(P5)}}$ becomes an LP with variables $\{Q_i\}$. It is easy to show that the complexity of the ZF-based beamforming is given as $\mathcal{O}(M^3N^2)$.

\subsection{MRT-Based Energy Beamforming}
In this subsection, we design the energy beamforming vectors at the S-ET based on the MRT principle, i.e., $\mv{\omega}_i=\sqrt{Q_i}\frac{\mv{g}_i}{\|\mv{g}_i\|}$, where $Q_i$ denotes the transmit power allocated to the SC $i\in\mathcal N$. In this case, problem $\mathtt{(P1)}$ can be expressed as
\begin{align}
\mathtt{(P6)}:~~~~&\nonumber\\
 \max\limits_{\{Q_i\},\lambda> 0} &\sum\limits_{i\in\mathcal N} \bigg(\bigg(\lambda-\frac{|{\mv{f}^H_i}{\mv{g}_i}|^2Q_i/\|\mv{g}_i\|^2+\sigma^2}{h_i}\bigg)^+{\varphi_i}\nonumber\\
&~~~~~~~~~+\|{\mv{g}_i}\|^2Q_i\bigg)\label{09251604}\\
\mathrm{s.t.}~~ &\sum\limits_{i\in\mathcal N} \left|{\mv{f}^H_i}{\mv{g}_i}\right|^2Q_i/\|\mv{g}_i\|^2 \le \Gamma\\
   &\sum\limits_{i\in\mathcal N} Q_i\leq Q_{\mathrm{sum}}\\
   &0\leq Q_i \leq Q_{\mathrm{peak}},~\forall i\in \mathcal{N}\\
   &\sum\limits_{i\in\mathcal N} \bigg(\lambda-\frac{|{\mv{f}^H_i}{\mv{g}_i}|^2Q_i/\|\mv{g}_i\|^2+\sigma^2}{h_i}\bigg)^+= P_{\mathrm{sum}}.
\end{align}
Problem $\mathtt{(P6)}$ can be similarly solved as for problem ${\mathtt{(P1)}}$. First, we solve for $\{Q_i\}$ under given $\lambda$, by using an algorithm similarly as Algorithm 2. Next, we use a one-dimensional search over $\lambda$ within its feasible region. For brevity, we omit the detailed algorithm here. Similarly as ${\mathtt{(P1)}}$ (but without SDP required), the computational complexity of the MRT-based energy beamforming
 is $\mathcal{O}(MN\log(1/\varepsilon_1)\log(1/\varepsilon_2)+M^3N\log ({1/\varepsilon_1})/\varepsilon_4)$.
\subsection{Conventional Energy Beamforming Design without Considering WIT System's Reaction}
In this subsection, we consider the conventional energy beamforming design without considering the primary WIT system's reaction. In this case, the received power maximization problem is formulated as
\begin{align}
\mathtt{(P7)}:\max\limits_{\{\mv{\omega}_i\}} ~~&\sum\limits_{i\in\mathcal N} {|{\mv{g}^H_i}\mv{\omega}_i|^2}\label{1809021741}\\
 \mathrm{s.t.} &\sum\limits_{i\in\mathcal N} |{\mv{f}^H_i}\mv{\omega}_i|^2 \le \Gamma\label{1809021742}\\
   &\sum\limits_{i\in\mathcal N} \|\mv{\omega}_i\|^2 \leq Q_{\mathrm{sum}}\label{1809021743}\\
   &\|\mv{\omega}_i\|^2 \leq Q_{\mathrm{peak}},\forall i\in \mathcal{N}.
   \end{align}
Problem ${\mathtt{(P7)}}$ is a non-convex QCQP, for which we can use the SDR to obtain the optimal solution, with the details omitted for brevity. Note that, in practice, the S-ER can still receive the power sent from the P-IT even without considering its reaction. Therefore, by letting $\{\bar{\mv{\omega}}_i\}$ denote the obtained optimal solution to ${\mathtt{(P7)}}$, the total power received by the S-ER is $\sum_{i\in\mathcal N} ({P_i^*(\bar{\mv{\omega}}_i,\bar\lambda){\varphi_i}+{|{\mv{g}^H_i}\mv{\omega}_i|}^2})$, in which $\bar\lambda$ is obtained based on (\ref{equ:ECS5}) under $\{\bar{\mv{\omega}}_i\}$. The computational complexity for solving SDP is $\mathcal{O}(\log ({1/\varepsilon_3}))$.
\section{Numerical Results}\label{sec:Results}


In this section, we provide numerical results to validate the performance of our proposed designs. In the simulation, we consider the path loss model $\chi(d/d_0)^{-\kappa}$, where $\chi$ denotes the channel power gain at a reference distance of $d_0$, $d$ denotes the transmission distance, and $\kappa$ denotes the path-loss exponent. For the channel vectors from the S-ET to the S-ER and the P-IR, we consider that each element is an independent CSCG random variable with mean zero and variance specified based on the above path loss. Furthermore, we consider that the S-ET, P-IT, P-IR, and S-ER are located at a two-dimensional space. The x-y coordinates of these nodes and other simulation parameters are summarized in Table $\mathrm{\uppercase\expandafter{\romannumeral1}}$, unless otherwise stated.

\begin{table}[H]
\centering
\caption{simulation parameters}
\begin{tabular}{|c|c|c|c|}
\hline
Parameter&Value&Parameter&Value\\
\hline
Location of S-ET&(0, 0)&$\sigma^2$&$10^{-9}$ W\\
\hline
Location of S-ER&(0, 5 m)&$\chi$&$-30$ dB\\
\hline
Location of P-IT&(0, 2.5 m)&$d_0$&1 m\\
\hline
Location of P-IR&(5 m, 0)&$\kappa$&3\\
\hline
$P_{\mathrm{sum}}$&3.2 W&$M$&4\\
\hline
$Q_{\mathrm{sum}}$&6.4 W&$\Gamma$&$1.28 \times 10^{-6}$ W\\
\hline
$Q_{\mathrm{peak}}$&0.1 W&$N$&64\\
\hline

\end{tabular}
\end{table}

\begin{figure}[htbp]
\center
\includegraphics[width=8cm,height=6cm]{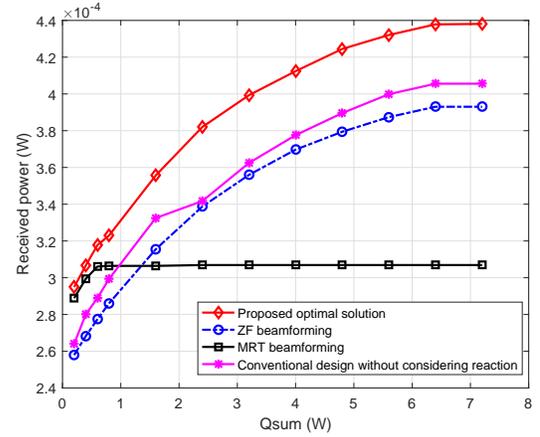}
\caption{The received power at the S-ER versus the maximum sum transmit power $Q_{\mathrm{sum}}$ at the S-ET.}\label{fig2}
\vspace{-1em}	
\end{figure}

Fig. \ref{fig2} shows the received power at the S-ER versus the maximum sum transmit power $Q_{\mathrm{sum}}$ at the S-ET. It is observed that our proposed design in Section \ref{sec:Solution} achieves the highest received power among the four schemes over the whole regime of $Q_{\mathrm{sum}}$. This shows the effectiveness of the proposed algorithm in maximizing the received total power by balancing between the power directly transferred from the S-ET and that from the P-IT. When $Q_{\mathrm{sum}}$ is small (e.g., less than 0.8 W), the MRT-based beamforming design outperforms both the ZF-based design and the conventional design without considering the primary WIT system's reaction. This is due to the fact that in this case, the interference power from the S-ET to the P-IR is small; as a result, the S-ET should set the beamforming directions towards the S-ER to maximize the directly transferred energy, and thus the MRT-based beamforming design is preferred. When $Q_\mathrm{sum}$ becomes large (e.g., more than $1.6$ W), the MRT-based beamforming design results in an unchanged received power, which is inferior to that achieved by the ZF-based design and the conventional design. This is because in this case, the interference power from the S-ET to the P-IR becomes large, and the MRT-based design cannot fully use the transmit power, thus leading to a limited wireless energy harvesting performance. By contrast, it is beneficial to use a ZF-like beamforming to minimize the interference towards the P-IR, so as to fully utilize the transmit power.

\begin{figure}[htbp]
\center
\includegraphics[width=8cm,height=6cm]{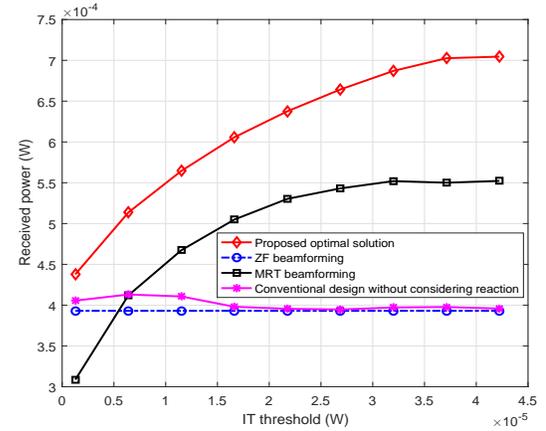}
\caption{The received power at the S-ER versus the IT threshold $\Gamma$.}\label{fig3}
\vspace{-1em}
\end{figure}
Fig. \ref{fig3} shows the received power at the S-ER versus the IT threshold $\Gamma$. It is observed that as $\Gamma$ increases, the received power by our proposed design and the MRT-based beamforming design increases, as the S-ET has more degrees of freedom in optimizing the transmit beamforming vectors or transmit power for improving the wireless energy harvesting performance. By contrast, as $\Gamma$ changes, the ZF-based beamforming design leads to unchanged received power at the S-ER. This is due to the fact that the ZF-based beamforming design is irrespective of $\Gamma$ (i.e., $\sum_{i\in\mathcal N} |{\mv{f}^H_i}\mv{\omega}_i|^2=0$). It is also observed that for the conventional design, the achieved received power at the S-ER fluctuates as the IT threshold becomes large. This is due to the fact that the conventional energy beamforming design does not consider the water-filling power allocation reaction at the primary WIT system, and thus leads to an uncontrolled power allocation at the P-IT, thus leading to fluctuated harvested energy at the S-ER. In addition, it is observed that the MRT-based beamforming design outperforms the ZF-based beamforming design and the conventional design when $\Gamma$ is large (e.g., $\Gamma\ge 11.52\times 10^{-6}$ W), but the reverse is true when $\Gamma$ is small (e.g., $\Gamma = 1.28\times 10^{-6}$ W). This indicates that the exploitation of the primary WIT system's reaction is more beneficial when the IT threshold $\Gamma$ becomes large.

\begin{figure}[htbp]
\center
\includegraphics[width=8cm,height=6cm]{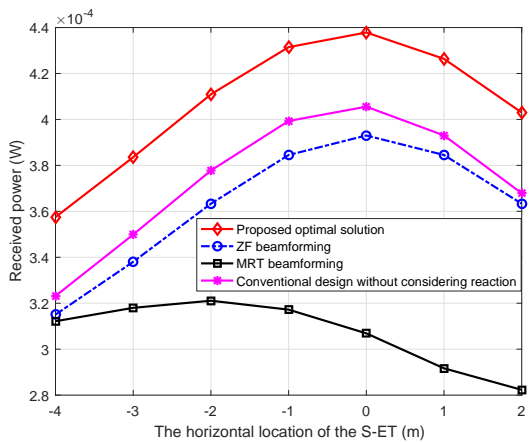}
\caption{The received power at the S-ER versus the horizontal location of S-ET.}\label{fig4}
\vspace{-2em}
\end{figure}

Fig. \ref{fig4} shows the received power at the S-ER versus the x-axis of the S-ET's location. It is observed that for all the four schemes, the received power at the S-ER first increases and then decreases. This is generally due the fact that as the x-axis of the S-ET changes, the distance between the S-ET and the S-ER also decreases and then increases. It is also observed that our proposed design significantly outperforms the other three benchmark schemes.

\begin{figure}[htbp]
\center
\includegraphics[width=8cm,height=6cm]{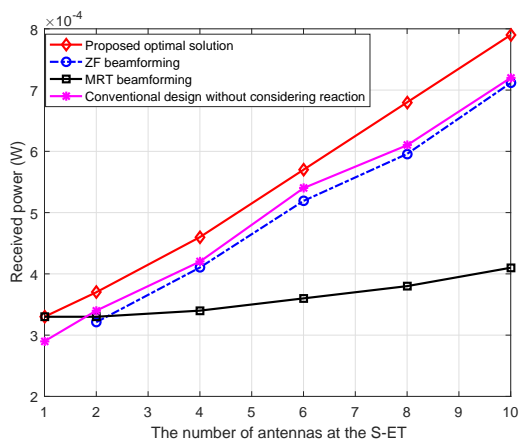}
\caption{The received power at the S-ER versus the number of antennas at the S-ET.}\label{fig5}
\vspace{-1em}
\end{figure}

Fig. \ref{fig5} shows the received power at the S-ER versus number of antennas at the S-ET $M$. Notice that the ZF beamforming design is only feasible when $M \ge 2$, so the curve for ZF beamforming starts from $M=2$. It is observed that, for all the four schemes, as $M$ increases, the received power at the S-ER becomes larger, due to the exploitation of the multi-antenna gain. It is also observed that our proposed design outperforms the three benchmark schemes for all values of $M$, and the performance gain over the MRT beamforming becomes more significant as $M$ increases.


\section{Concluding Remark}\label{sec:Conclusion}
This paper studies a cognitive or secondary multi-antenna WPT system over a multi-carrier channel, in which we exploit the primary WIT system's water-filling power allocation for improving the wireless energy harvesting performance. In particular, we design the S-ET's transmit energy beamforming over SCs to maximize the total energy received at the S-ER from both the S-ET and the P-IT, subject to the S-ET's maximum transmit power constraint, and the maximum interference power constraint imposed at the P-IR to protect the primary WIT. We propose an efficient algorithm to obtain a high-quality solution to this non-convex optimization problem. The proposed beamforming design can efficiently balance between the direct energy transfer from the S-ET to the S-ER, versus the reactive energy transfer from the P-IT (that is controlled by the S-ET by purposely designed one-way interference), thus leading to improved performance as compared to conventional designs without considering such reaction at the primary system. It is our hope that the proposed approaches can provide new insights on the design of coexisting WPT and WIT systems in next-generation self-sustainable IoT networks. Due to space limitation, there are several important issues that are not addressed in this paper. We briefly discuss these issues in the following to motivate future work.
\begin{itemize}

\item The implementation of the proposed optimization algorithm requires the S-ET to perfectly know the CSI of $\{h_i\}$, $\{\psi_i\}$, $\{\mv{g}_i\}$, and $\{\mv{f}_i\}$. Though generally difficult, such CSI acquisition is feasible in practice, as explained in detail as follows. First, in order for the S-ET to obtain $\{\mv{g}_i\}$ and $\{\psi_i\}$, the S-ER can overhear the transmitted signals (particularly pilots) from the S-ET and the P-IT, perform channel estimation, quantize the estimated CSI, and then send the quantized version of $\{\mv{g}_i\}$ and $\{\psi_i\}$ to the S-ET (see \cite{Love} for an overview on the limited feedback technique in wireless communications). Next, the S-ET can also acquire the CSI of $\{\mv{f}_i\}$ and $\{h_i\}$ by monitoring the potential signaling channel from the P-IR to the P-IT. On one hand, the S-ET can overhear the pilot signals sent from the P-IR to estimate the reverse link of $\{\mv{f}_i\}$, which can then be used as an approximation of $\{\mv{f}_i\}$.{\footnote{If the signaling channel from the P-IR to the P-IT is implemented over the same frequency band as that for WPT, then the approximation is accurate due to the channel reciprocity.}} On the other hand, in order to get an estimation of $\{h_i\}$, the S-ET can intercept the channel quality indicators (CQIs) for different SCs, which are sent from the P-IR to the P-IT to facilitate the water-filling power allocation for the primary WIT system.{\footnote{Notice that similar CSI acquisition techniques have been used in the wireless surveillance literature (see, e.g., \cite{duan,XD} and the references therein) to acquire the CSI of another wireless communication system.}} Nevertheless, notice that with the above procedures performed, the S-ET may only know imperfect CSI of $\{h_i\}$, $\{\psi_i\}$, $\{\mv{g}_i\}$, and $\{\mv{f}_i\}$ due to, e.g., channel estimation errors, channel quantization errors, and the imperfect channel reciprocity between the forward and reverse links. Therefore, it is an important issue to properly design the CSI acquisition under our setup to balance between the acquisition overhead and the CSI accuracy. Furthermore, how to optimize the energy beamforming over SCs under such imperfect CSI (e.g., via robust optimization techniques \cite{LUO}) is another issue that requires further investigation.

\item This paper considered the case with a single antenna at the S-ER. In practice, the S-ER can also deploy multiple antennas to improve the energy harvesting performance. In this case, the energy beamforming design critically depends on the energy harvesting (or rectifier) architecture of the S-ER. For instance, the S-ER can use one rectifier for all antennas, multiple rectifiers each for an antenna, or even more generic rectifier architecture (with adaptive power splitting) for energy harvesting \cite{Ma}. In this case, depending on the used energy harvesting architecture at the multi-antenna S-ER,  how to optimize the transmit energy beamforming at the S-ET, together with the energy harvesting processing, is an interesting problem for future study. Furthermore, with multiple rectifiers and multiple antennas, the nonlinear RF-to-DC conversion must be taken into account, which will make such problems more difficult.

\item Although this paper considered a single S-ET with multiple transmit antennas co-located, our design can be extended to the  distributed energy beamforming scenario with multiple S-ETs cooperatively sending  energy signals to a single-antenna S-ER. In this case, new individual transmit power constraints each for one S-ET should be taken into account (see, e.g., \cite{Zhang}), which are generally more difficult to be dealt with. Furthermore, how to design distributed algorithms under only local CSI at each S-ET is a very challenging problem to be studied.

\item This paper considered that both the P-IT and P-IR are deployed with a single antenna in the primary WIT system. If the P-IT and/or P-IR are deployed with multiple antennas, then beamforming/precoding design should be employed together with the (water-filling) power allocation to maximize the communication rate in the primary WIT system. However, the joint beamforming and power allocation at the P-IT/P-IR critically depend on the availability of the multi-antenna wireless channels that are generally vectors or even matrices. How to obtain such information and exploit the {\it beamforming} reaction in the primary WIT system is an interesting problem that is more difficult to be solved optimally.

\end{itemize}


\begin{thebibliography}{1}
\bibliographystyle{IEEEbib}
\bibitem{1}
S. Bi, C. K. Ho, and R. Zhang, ``Wireless powered communication: Opportunities and challenges,'' {\it IEEE Commun. Mag.}, vol. 53, no. 4, pp. 117--125, Apr. 2015.
\bibitem{2}
K. Huang and V. K. N. Lau, ``Enabling wireless power transfer in cellular networks: Architecture, modeling and deployment,'' {\it IEEE Trans. Wireless Commun.}, vol. 13, no. 2,
pp. 902--912, Feb. 2014.
\bibitem{Xia}
M. Xia and S. Aissa, ``On the efficiency of far-field wireless power transfer," {\it IEEE Trans. Signal Process.}, vol. 63, no. 11, pp. 2835--2847, Jun. 2015.
\bibitem{12}
R. Zhang and C. K. Ho, ``MIMO broadcasting for simultaneous wireless
information and power transfer,'' {\it IEEE Trans. Wireless Commun.}, vol. 12,
no. 5, pp. 1989--2001, May 2013.
\bibitem{15}
X. Zhou, R. Zhang, and C. K. Ho, ``Wireless information and power transfer:
Architecture design and rate-energy tradeoff,'' {\it IEEE Trans. Commun.},
vol. 61, no. 11, pp. 4757--4767, Nov. 2013.
\bibitem{16}
J. Xu, L. Liu, and R. Zhang, ``Multiuser MISO beamforming for simultaneous
wireless information and power transfer,'' {\it IEEE Trans. Signal Process.}, vol. 62, no. 3,
pp. 4798--4810, Sep. 2014.
\bibitem{Costa}
Y. Chen, D. B. da Costa, and H. Ding, ``Interference analysis in wireless power transfer," {\it IEEE Commun. Lett.}, vol. 21, no. 10, pp. 2318--2321, Oct. 2017.
\bibitem{20}
S. Bi, Y. Zeng, and R. Zhang, ``Wireless powered communication
networks: An overview,'' {\it IEEE Wireless Commun.}, vol. 23, no. 2,
pp. 10--18, Apr. 2016.
\bibitem{22}
L. Liu, R. Zhang, and K. Chua, ``Multi-antenna wireless powered communication
with energy beamforming,'' {\it IEEE Trans. Commun.}, vol. 62, no. 12, pp. 4349--4361, Dec. 2014.
\bibitem{222}
Y. Che, J. Xu, L. Duan, and R. Zhang, ``Multiantenna wireless powered communication with cochannel energy and information transfer,'' {\it IEEE Commun. Lett.}, vol. 19, no. 12, pp. 2266--2269, Dec. 2015.
\bibitem{you}
C. You, K. Huang, and H. Chae, ``Energy efficient mobile cloud computing powered by wireless energy transfer," {\it IEEE J. Sel. Areas Commun.}, vol. 34, no. 5, pp. 1757--1771, May 2016.
\bibitem{222222}
F. Wang, J. Xu, X. Wang, and S. Cui, ``Joint offloading and computing optimization in wireless powered mobile-edge computing systems," {\it IEEE Trans. Wireless Commun.}, vol. 17, no. 3, pp. 1784--1797, Mar. 2018.
\bibitem{Bi}
S. Bi and Y. J. Zhang, ``Computation rate maximization for wireless powered mobile-edge computing with binary computation offloading," {\it IEEE Trans. Wireless Commun.}, vol. 17, no. 6, pp. 4177--4190, Jun. 2018.
\bibitem{511}
B. Clerckx and E. Bayguzina, ``Waveform design for wireless power transfer," {\it IEEE Trans. Signal Process.}, vol. 64, no. 23, pp. 6313--6328, Dec. 2016.
\bibitem{111}
M. R. V. Moghadam, Y. Zeng and R. Zhang, ``Waveform optimization for radio-frequency wireless power transfer,"  in {\it Proc. IEEE SPAWC}, 2017, pp. 1--6.
\bibitem{Zeng}
Y. Zeng and R. Zhang, ``Optimized training for net energy maximization in multi-antenna wireless energy transfer over frequency-selective channel," {\it IEEE Trans. Commun.}, vol. 63, no. 6, pp. 2360--2373, Jun. 2015.
\bibitem{51}
X. Zhou, R. Zhang, and C. K. Ho, ``Wireless information and power transfer in multiuser OFDM systems," {\it IEEE Trans. Wireless Commun.}, vol. 13, no. 4, pp. 2282--2294, Apr. 2014.
\bibitem{Huang}
Y. Huang and B. Clerckx, ``Large-scale multiantenna multisine wireless power transfer,"  {\it IEEE Trans. Signal Process.}, vol. 65, no. 21, pp. 5812--5827, Nov. 2017.
\bibitem{5}
S. Lee and R. Zhang, ``Distributed wireless power transfer with energy feedback,'' {\it IEEE Trans. Signal Process.}, vol. 65, no. 7, pp. 1685--1699, Apr. 2017.
\bibitem{3}
J. Xu and R. Zhang, ``Energy beamforming with one-bit feedback,'' {\it IEEE Trans. Signal Process.}, vol. 62, no. 20, pp. 5370--5381, Oct. 2014.
\bibitem{4}
J. Xu and R. Zhang, ``A general design framework for MIMO wireless energy transfer with limited feedback,'' {\it IEEE Trans. Signal Process.}, vol. 64, no. 20, pp. 2475--2488, May 2016.
\bibitem{11}
Y. Zeng, B. Clerckx, and R. Zhang, ``Communications and signals design for wireless power transmission,'' {\it IEEE Trans. Commun.}, vol. 65, no.5, pp. 2264--2290, May 2017.
\bibitem{Goldsmith}
A. Goldsmith, S. A. Jafar, I. Maric, and S. Srinivasa, ``Breaking spectrum gridlock with cognitive radios: An information theoretic perspective," {\it Proc.  IEEE}, vol. 97, no. 5, pp. 894--914, May 2009.
\bibitem{hay}
S. Haykin, ``Cognitive radio: Brain-empowered wireless communications", {\it IEEE J. Sel. Areas Commun.}, vol. 23, pp. 201--220, Feb. 2005.
\bibitem{sad}
Q. Zhao and B. M. Sadler, ``A survey of dynamic spectrum access", {\it IEEE Signal Process. Mag.}, pp. 79--89, May 2007.
\bibitem{g}
I. Maric, A. Goldsmith, G. Kramer, and S. Shamai (Shitz), ``On the capacity of interference channels with a partiallycognitive
transmitter", {\it 2007 IEEE International Symposium On Information Theory}, Nice, France, Jun. 2007.
\bibitem{32}
X. Luo, Y. Zeng, and R. Zhang, ``Cognitive wireless power transfer with information helping,''  {\it IEEE Wireless Commun. Lett.}, vol. 6, no. 3, pp. 346--349, Jun. 2017.
\bibitem{322}
J. Xu, S. Bi, and R. Zhang, ``Multiuser MIMO wireless energy transfer with coexisting opportunistic communication,'' {\it IEEE Wireless Commun. Lett.}, vol. 4, no 3, pp. 273--276, Jun. 2015.
\bibitem{29}
F. Yao, H. Wu, Y. Chen, Y. Liu, and T. Liang, ``Cluster-based collaborative spectrum sensing for energy harvesting cognitive wireless communication network,'' {\it IEEE Access}, vol. 5, pp. 9266--9276, May 2017.
\bibitem{26}
Z. Yang, Z. Ding, P. Fan, and G. K. Karagiannidis, ``Outage performance
of cognitive relay networks with wireless information and power
transfer,'' {\it IEEE Trans. Veh. Technol.}, vol. 65, no. 5, pp. 3828--3833,
May 2016.
\bibitem{25}
Z. Wang, Z. Chen, B. Xia, L. Luo, and J. Zhou, ``Outage analysis of cognitive relay networks with
energy harvesting and information transfer,'' in {\it Proc. IEEE Int. Conf.
Commun. (ICC)}, Sydney, NSW, Australia, Jun. 2014, pp. 4348--4353.
\bibitem{23}
S. Lee and R. Zhang, ``Cognitive wireless powered network: Spectrum
sharing models and throughput maximization,'' {\it IEEE Trans. Cogn.
Commun. Netw.}, vol. 1, no. 3, pp. 335--346, Sep. 2015.
\bibitem{31}
Z. Qin, Y. Liu, Y. Gao, M. Elkashlan, and A. Nallanathan, ``Wireless powered cognitive radio networks with compressive sensing and matrix completion,'' {\it IEEE Trans. Commun.}, vol. 65, no. 4, pp. 1464--1476, Apr. 2017.
\bibitem{30}
X. Ji, J. Xu, Y. L. Che, Z, Fei, and R. Zhang, ``Adaptive mode switching for cognitive wireless powered communication systems,'' {\it IEEE Wireless Commun. Lett.}, vol. 6, no. 3, pp. 386--389, Apr. 2017.
\bibitem{24}
S. A. Mousavifar, Y. Liu, C. Leung, M. Elkashlan, and T. Q. Duong,
``Wireless energy harvesting and spectrum sharing in cognitive radio,''
in {\it Proc. IEEE Veh. Technol. Conf. (VTC Fall)}, Vancouver, BC, Canada,
Sep. 2014, pp. 1--5.
\bibitem{27}
C. Xu, M. Zheng, W. Liang, H. Yu, and Y.-C. Liang, ``Outage
performance of underlay multihop cognitive relay networks with energy
harvesting,'' {\it IEEE Commun. Lett.}, vol. 20, no. 6, pp. 1148--1151,
Jun. 2016.
\bibitem{28}
S. Yin, Z. Qu, Z. Wang, and L. Li, ``Energy-efficient cooperation in cognitive wireless powered networks,'' {\it IEEE Commun. Lett.}, vol. 21, no. 1, pp. 128--131,  Jan. 2017.



\bibitem{37}
S. Boyd and L. Vandenberghe, {\it Convex Optimization}, Cambidge Univ.
Press, 2004.
\bibitem{371}
B. Clerckx, R. Zhang, R. Schober, D. W. K. Ng, D. I. Kim, and H. V. Poor, ``Fundamentals of wireless information and power transfer: From RF energy harvester models to signal and system designs,'' {\it IEEE J. Sel. Areas Commun.}, vol. 37, no. 1, pp. 4-33, Jan. 2019.


\bibitem{non-linearModel}
E. Boshkovska, D. W. K. Ng, N. Zlatanov, and R. Schober, ``Practical non-linear energy harvesting model and resource allocation for SWIPT systems," {\it IEEE Commun. Lett.}, vol. 19, no. 12, pp. 2082--2085, Dec. 2015.









\bibitem{34}
W. Yu and R. Lui, ``Dual methods for nonconvex spectrum
optimization of multicarrier systems,'' {\it IEEE Trans. Commun.}, vol. 54, no. 7, pp. 1310--1322, Jul. 2006.
\bibitem{duan}
J. Xu, L. Duan, and R. Zhang, ``Proactive eavesdropping via cognitive jamming in fading channels," {\it IEEE Trans. Wireless Commun.}, vol. 16, no. 5, pp. 2790--2806, May 2017.
\bibitem{35}
Z. Q. Luo, W. K. Ma, A. M. C. So, Y. Ye, and S. Zhang, ``Semidefinite relaxation of quadratic optimization problems," {\it IEEE Signal Process. Mag.}, vol 27, no. 3, pp. 20--34, Apr. 2010.

\bibitem{36}
S. Boyd, {\it Ellipsoid method}, Stanford, CA, USA. [Online]. Available:\url{https://web.stanford.edu/class/ee364b/lectures/ellipsoid_method_slides.pdf}


\bibitem{com}
S. Boyd, {\it Interior-point methods}, Stanford, CA, USA. [Online]. Available:\url{https://web.stanford.edu/class/ee364a/lectures/barrier.pdf}

\bibitem{Love}
D. J. Love, R. W. Heath, V. K. N. Lau, D. Gesbert, B. D. Rao and M. Andrews, ``An overview of limited feedback in wireless communication systems," {\it IEEE J. Sel. Areas Commun.}, vol. 26, no. 8, pp. 1341--1365, Oct. 2008.
\bibitem{XD}
J. Xu, L. Duan, and R. Zhang, ``Surveillance and intervention of infrastructure-free mobile communications: a new wireless security paradigm," {\it IEEE Wireless Commun.}, vol. 24, no. 4, pp. 152--159, Aug. 2017.

\bibitem{LUO}
S. A. Vorobyov, A. B. Gershman and Zhi-Quan Luo, ``Robust adaptive beamforming using worst-case performance optimization: a solution to the signal mismatch problem," {\it IEEE Trans. Signal Process.}, vol. 51, no. 2, pp. 313--324, Feb. 2003.

\bibitem{Ma}
G. Ma, J. Xu, Y. Zeng and M. R. V. Moghadam, ``A generic receiver architecture for MIMO wireless power transfer with nonlinear energy harvesting," {\it IEEE Signal Process. Lett.}, vol. 26, no. 2, pp. 312--316, Feb. 2019.

\bibitem{Zhang}
R. Zhang, ``Cooperative multi-cell block diagonalization with per-base-station power constraints," {\it IEEE J. Sel. Areas Commun.}, vol. 28, no. 9, pp. 1435--1445, Dec. 2010.
\end{thebibliography}
\end{document}